\documentclass[aps,prl,superscriptaddress,showpacs,twocolumn]{revtex4}
\usepackage{amsmath}
\usepackage{amssymb}
\usepackage{graphicx}
\usepackage{dcolumn}

\begin{document}

\title{Observation of universal behaviour of ultracold quantum critical gases%
}
\author{Hongwei Xiong}
%\email{xionghongwei@wipm.ac.cn}
\affiliation{State Key Laboratory of Magnetic Resonance and Atomic and Molecular Physics,
Wuhan Institute of Physics and Mathematics, Chinese Academy of Sciences,
Wuhan 430071, P. R. China}
\author{Xinzhou Tan}
\affiliation{State Key Laboratory of Magnetic Resonance and Atomic and Molecular Physics,
Wuhan Institute of Physics and Mathematics, Chinese Academy of Sciences,
Wuhan 430071, P. R. China}
\affiliation{Graduate School of the Chinese Academy of Sciences, P. R. China}
\author{Bing Wang}
\affiliation{State Key Laboratory of Magnetic Resonance and Atomic and Molecular Physics,
Wuhan Institute of Physics and Mathematics, Chinese Academy of Sciences,
Wuhan 430071, P. R. China}
\affiliation{Graduate School of the Chinese Academy of Sciences, P. R. China}
\author{Lijuan Cao}
\affiliation{State Key Laboratory of Magnetic Resonance and Atomic and Molecular Physics,
Wuhan Institute of Physics and Mathematics, Chinese Academy of Sciences,
Wuhan 430071, P. R. China}
\affiliation{Graduate School of the Chinese Academy of Sciences, P. R. China}
\author{Baolong L\H{u}}
%\email{baolong _lu@wipm.ac.cn}
\affiliation{State Key Laboratory of
Magnetic Resonance and Atomic and Molecular Physics, Wuhan Institute
of Physics and Mathematics, Chinese Academy of Sciences, Wuhan
430071, P. R. China}
\date{\today }

\begin{abstract}
Quantum critical matter has already been studied in many systems, including
cold atomic gases. We report the observation of a universal behaviour of
ultracold quantum critical Bose gases in a one-dimensional optical lattice.
In the quantum critical region above the Berezinskii-Kosterlitz-Thouless
transition, the relative phase fluctuations between neighboring
subcondensates and spatial phase fluctuations of quasi-2D subcondensates
coexist. We study the density probability distribution function when both
these two phase fluctuations are considered. A universal exponential density
probability distribution is demonstrated experimentally, which agrees well
with a simple theoretical model by considering these two phase fluctuations.

%PACS: 03.75.Lm; 03.75.Kk; 05.30.Jp
\end{abstract}

\maketitle

The nature of quantum criticality \cite{Hertz,Sachdev,Cole} driven by
quantum fluctuations is still a great puzzle, despite of the remarkable
advances in heavy-fermion metals and rare-earth-based intermetallic
compounds, etc \cite{Gegen}. New understanding of quantum criticality is
widely believed to be a key to resolving open questions in metal-insulator
transitions \cite{Imada}, high temperature superconductivity \cite{HS} and
novel material design, etc. Cold atoms in optical lattices provide a unique
chance to not only simulate other strongly correlated systems \cite%
{RMP-Bloch}, but also study some models unaccessible in solid state systems,
particularly for the Bose-Hubbard model \cite{Fisher,Jaksch,Bloch}. Despite
of its complexity, a strongly correlated system in quantum critical regime
is expected to exhibit a universal behaviour described by a certain physical
quantity.

Here we report the observation of universal behaviour for ultracold quantum
critical Bose gases in a one-dimensional optical lattice. Density
probability distributions of the released gases are measured for different
depths of the lattice potential. It was found that the density probability
follows a simple exponential law when the Bose gases reach the quantum
critical region above the Berezinskii-Kosterlitz-Thouless (BKT) transition
\cite{BKTtheory1,BKTtheory2,Pol,BKTexp}. This universal behaviour can be
well understood in terms of our theoretical model considering both the
relative phase fluctuations of quasi-2D subcondensates and spatial phase
fluctuations of individual subcondensates above the BKT transition. The
method of density probability distribution should provide a unique tool for
identifying certain quantum phases of optical lattice systems.

Ultracold Bose atoms in 1D, 2D and 3D optical lattices are widely studied by
the Bose-Hubbard model \cite{Fisher}. %At zero temperature, there is a
%continuous superfluid-Mott insulator quantum phase transition. At finite
%temperatures, there exists a quantum critical (QC) region between the
%superfluid (SF) and Mott insulator (MI) regions \cite{Sachdev,Ho,Cap,Kato}.
At zero temperature, there is a continuous quantum phase transition from
superfluid (SF) to Mott insulator (MI). Because of the strongly correlated
quantum behaviour, the quantum critical point (QCP) at absolute zero
temperature distorts strongly the structure of the phase diagram at finite
temperatures, leading to the emergence of an unconventional `V-shaped'
quantum critical region \cite{Sachdev,Ho,Cap,Kato,Chin,Haz}(Fig. 1). In
analogy with a black hole, the crossover to quantum critical gas involves
crossing a `material event horizon', which implies strongly that the quantum
critical gas has a simple and universal behaviour \cite{Cole}.

\begin{figure}[tbp]
\centering
\includegraphics[width=0.7\linewidth,angle=270]{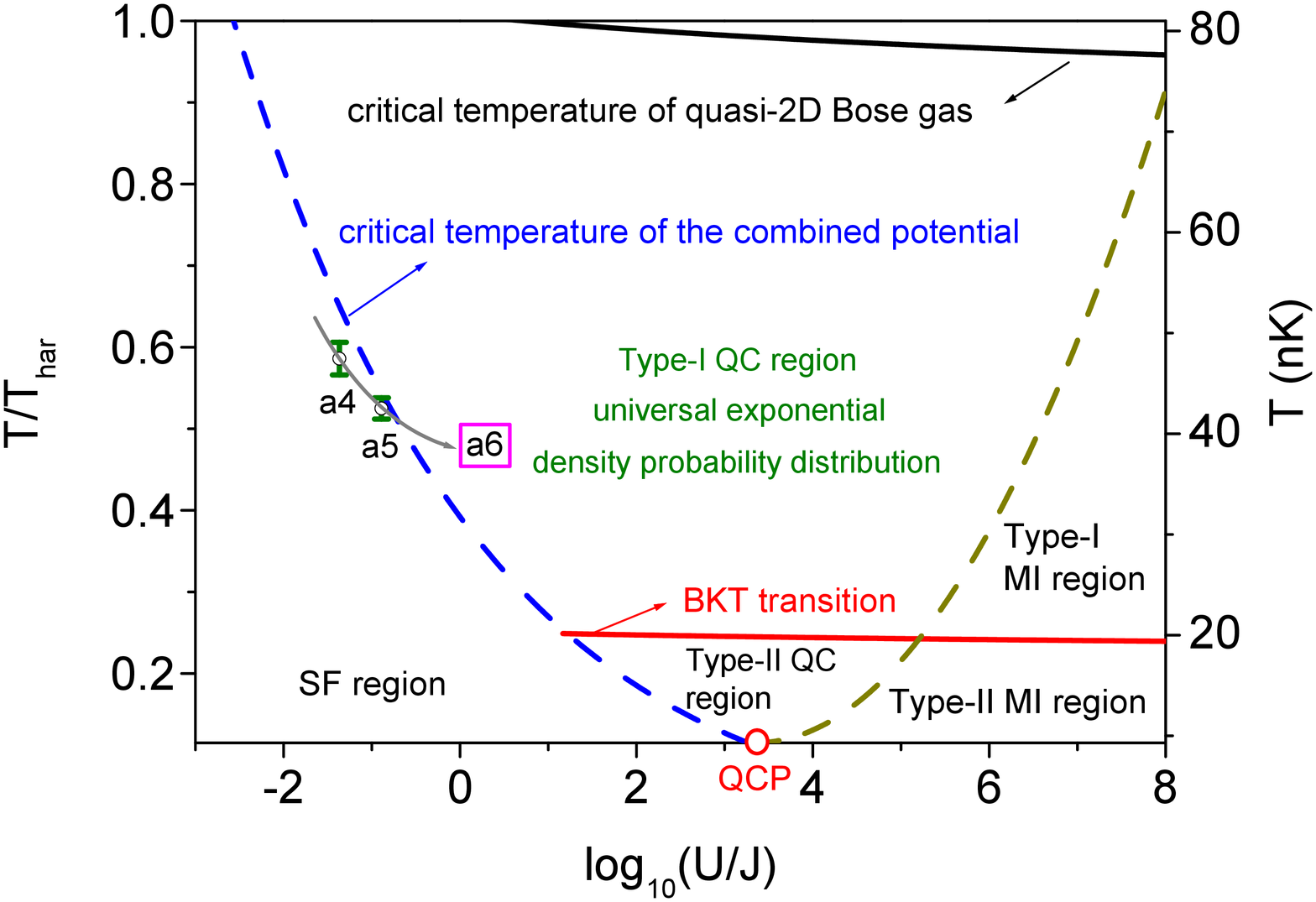}
\caption{Finite temperature phase diagram of ultracold Bose gases in a
one-dimensional optical lattice. $U$ and $J$ represent the strengths of the
on-site repulsion and of the nearest neighbor hopping in the Bose-Hubbard
model, respectively. By increasing the strength of an optical lattice, $U/J$
can be increased by several orders of magnitude. The existence of the
quantum critical point (QCP) at zero temperature leads to a `V-shaped'
quantum critical (QC) region. Apart from the ordinary SF-QC-MI transition,
the BKT transition for quasi-2D subcondensates can also occur at
sufficiently low temperatures. In the QC region above the BKT transition
(Type-I QC region), there exists a universal behaviour in the density
probability distribution. All the phase boundary lines, except for the
dashed line between QC and MI regions, are calculated with a total atom
number of $1.1\times 10^{5}$ and trap parameters used in the experiment. $%
T_{har}$ is the critical temperature for the purely harmonic trap in absence
of the optical lattice. The gray line displays a phase transition path from
SF to Type-I QC region. Data points of (a4)-(a6) along the gray line are
obtained from the corresponding absorption images in Fig. 2}
\end{figure}

In the commonly used 3D optical lattice systems, the atoms in a lattice site
are strongly confined in all directions, with a mean occupation number per
lattice site of about $1\sim 3$ \cite{Bloch}. Thus the spatial phase
fluctuations for the atoms in a single lattice site can be omitted, and the
relevant theoretical calculations based on the Bose-Hubbard model \cite%
{Fisher,Jaksch} and Wannier function \cite{Kohn} can give quantitative
descriptions of almost all experimental phenomena\cite{RMP-Bloch}. In
contrast, for ultracold Bose atoms in a 1D optical lattice, unique
characteristics may arise due to the following two reasons:

\textbf{(1) Quasi-2D Bose gas:} In an experiment of 1D optical lattice
system as ours, there can be hundreds of atoms in a lattice site. The Bose
gas in a lattice site becomes quasi-two-dimensional (quasi-2D) \cite%
{Pedri,Burger,Had}, if the local trapping frequency $\widetilde{\omega }_{z}$
of a lattice site in the lattice direction satisfies the condition $\hbar
\widetilde{\omega }_{z}>>k_{B}T$ \cite{Pethick}.

\textbf{(2) BKT transition:} For such a quasi-2D Bose gas, the critical
temperature in the occupied lattice site is given by $T_{2D}\approx \hbar
\omega _{\perp }\left( N_{l}\zeta \left( 2\right) \right) ^{1/2}/k_{B}$,
with $N_{l}$ being the atomic number in a lattice site. Well below $T_{2D}$,
the whole system becomes a chain of subcondensates. If there is no
correlation between the subcondensates in different lattice sites, a
quasi-2D gas undergoes a BKT transition at $T_{BKT}=T_{2D}/4$ \cite%
{BKTtheory1,BKTtheory2}. Beyond this critical value of $T_{BKT}$, it is
favorable to create vortices in quasi-2D subcondensates, and the unbinding
of bound vortices will lead to strong spatial phase fluctuations within each
subcondensate.

%\textit{Finally, based on the condition }$U/t>2.2\overline{n}$\textit{\ \cite%
%{Zwerger} of the SF-MI transition in 1D optical lattice at zero temperature,
%it is much more stringent to get true Mott insulator compared with that of
%3D optical lattice, because }$\overline{n}>>1$\textit{\ and much smaller }$U$%
%\textit{\ for 1D optical lattice with the same }$s$\textit{\ \cite{Had}.
%However, the quantum critical region can be reached much easier at finite
%temperatures, as shown in Fig. 1.%
%\textit{\ In the quantum critical region, we stress that although the whole
%system cannot be regarded as a superfluid due to completely random relative
%phases between neighboring lattice sites, there still exists quasicondensate
%if }$T<T_{2D}$\textit{.}

Since the BKT transition is crucial to revealing the property of cold atoms
in a 1D optical lattice, its critical temperature is also plotted in the
phase diagram (red line in Fig. 1). The BKT transition line divides the QC
region into two parts (Type I and Type II). Our experiments focus on the
transition from SF region to QC region above the BKT transition.

\begin{figure}[tbp]
\includegraphics[width=1.1\linewidth,angle=0]{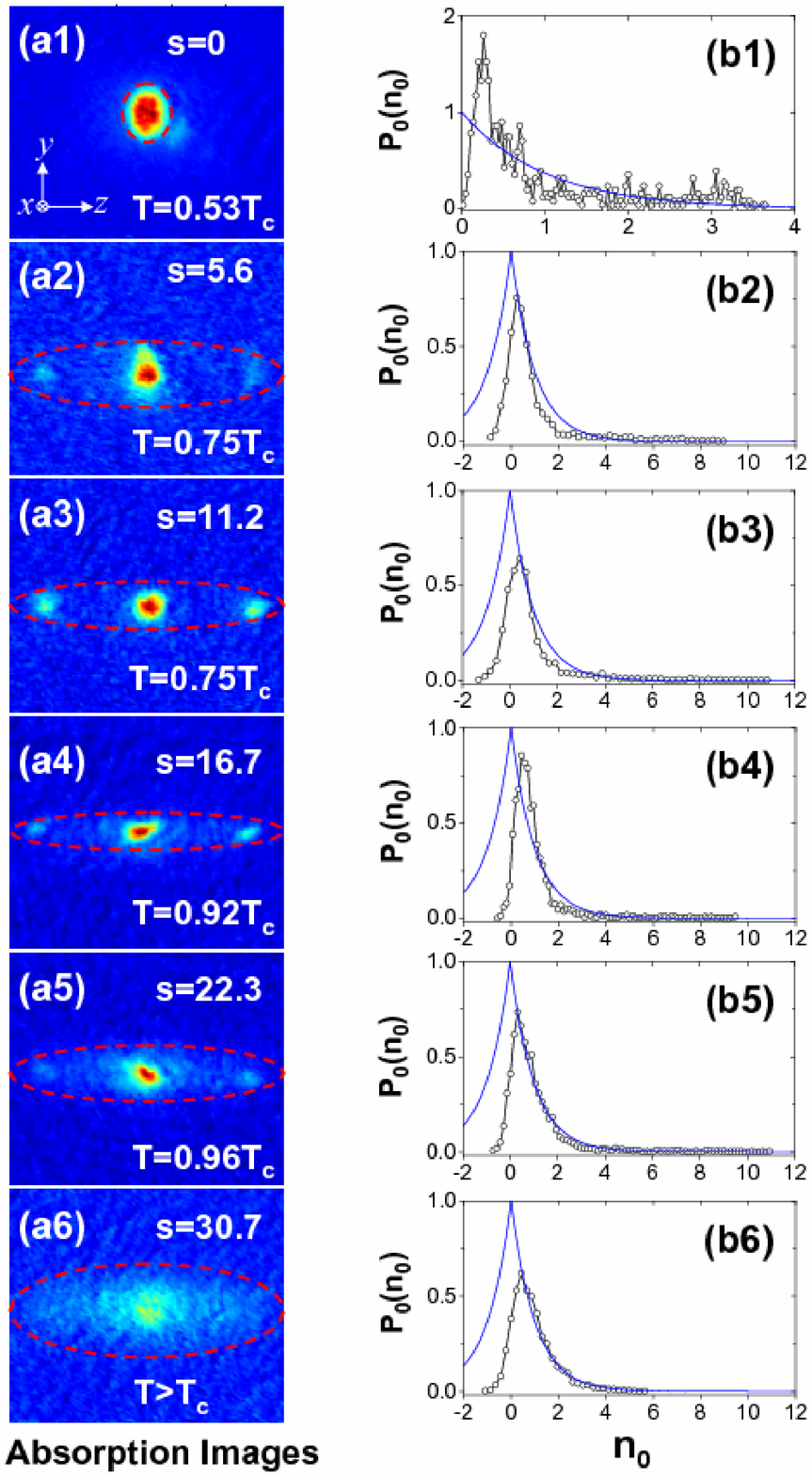}
\caption{Density probability distribution for different lattice depths. (%
\textbf{a1})-(\textbf{a6}) in the left column are absorption images showing
the density distribution of the released atomic clouds. The field of view is
$0.9\times 0.9$ \textrm{mm}$^{2}$, and the pixel size is $\Delta
^{2}=9.0\times 9.0$ $\mathrm{\protect\mu m}^{2}$. $T_{c}$ denotes the
critical temperature of the atomic gas in the combined trap. For moderate $s$%
, the interference fringes are prominent. Further increasing $s$, however,
we see complete disappearance of the interference fringes (\textbf{a6}).
This is due to the crossover from SF to QC region as $s$ is increased. (%
\textbf{b1})-(\textbf{b6}) in the right column are the corresponding density
probability distributions. Open circles are the data points calculated from
the density distributions of the images in the left column. The blue solid
curves give the exponential density probability distribution $P_{0}^{e}$
defined in text. When $s$ becomes large enough, the data points agree well
with the exponential curve (see \textbf{b5} and \textbf{b6}), which shows
clearly a universal behaviour in the QC region above the BKT transition. }
\end{figure}

The experiments started with pure $^{87}$Rb condensates confined in a
magnetic trap with axial and transverse trapping frequencies of $\left\{
\omega _{\bot },\omega _{z}\right\} =2\pi \left\{ 83.7,7.6\right\} $ \textrm{%
Hz}. The 1D optical lattice was formed by a retroreflected laser beam of $%
\lambda =800\,$\textrm{nm} along the axis ($z$ direction) of the condensate.
This laser beam was ramped up to a given intensity over a time of $50$
\textrm{ms}, yielding a lattice potential $V_{opt}=sE_{R}\sin ^{2}\left(
2\pi z/\lambda \right) $, with $E_{R}$ being the recoil energy of an atom
absorbing one lattice photon. After a holding time of $10$ \textrm{ms}, we
suddenly switched off the combined potential and allowed the cold atomic
cloud to expand freely for a time of $30$ \textrm{ms}.

The expanded atomic gases were probed using the conventional absorption
imaging technique. Since the probe beam is applied along $x$ direction, what
an absorption image records is the two-dimensional ($y-z$) column density
profile of the atomic cloud. We denote by $N_{1}^{ph}\left( y,z\right) $ and
$N_{2}^{ph}\left( y,z\right) $ the number of photons detected in the pixel
at position $\left( y,z\right) $ with and without the atomic cloud,
respectively. Then the density distribution is written as $n_{2D}\left(
y,z\right) =\ln \left[ N_{2}^{ph}\left( y,z\right) /N_{1}^{ph}\left(
y,z\right) \right] \Delta ^{2}/\sigma _{e}$, with $\sigma _{e}$ being the
absorption cross section of a single atom and $\Delta $ the pixel size.

The left column of Fig. 2 ((a1)-(a6)) displays the density distributions for
different lattice depths $s$. The temperatures of the system were inferred
from the condensate fractions as described in the Appendix. From the top
three images ((a1)-(a3)), we see an increasing of the side peaks in the
interference patterns as $s$ is increased. Further increasing $s$, we find
the interference fringes become blurred and disappear eventually (see
(a4)-(a6)), similar to the experimental observation in Ref. \cite{Orzel}.
The gradual disappearance of the interference fringes can be partially
explained by the increasing of the random relative phase \cite{Orzel} during
the SF-QC transition. In Figs. 2(a5)-(a6), except for the density
fluctuations along $z$ direction associated with the random relative phase
between different subcondensates, we also see significant density
fluctuations along $y$ direction. The simultaneous existence of density
fluctuations in these two directions is further demonstrated in Fig. 3 (c)
and (d) for a typical case of $s=30.7$. These experimental results suggest
that in the QC region above the BKT transition, both the random relative
phases and spatial phase fluctuations play important roles in the expanded
density distribution.

\begin{figure}[tbp]
\includegraphics[width=0.7\linewidth,angle=270]{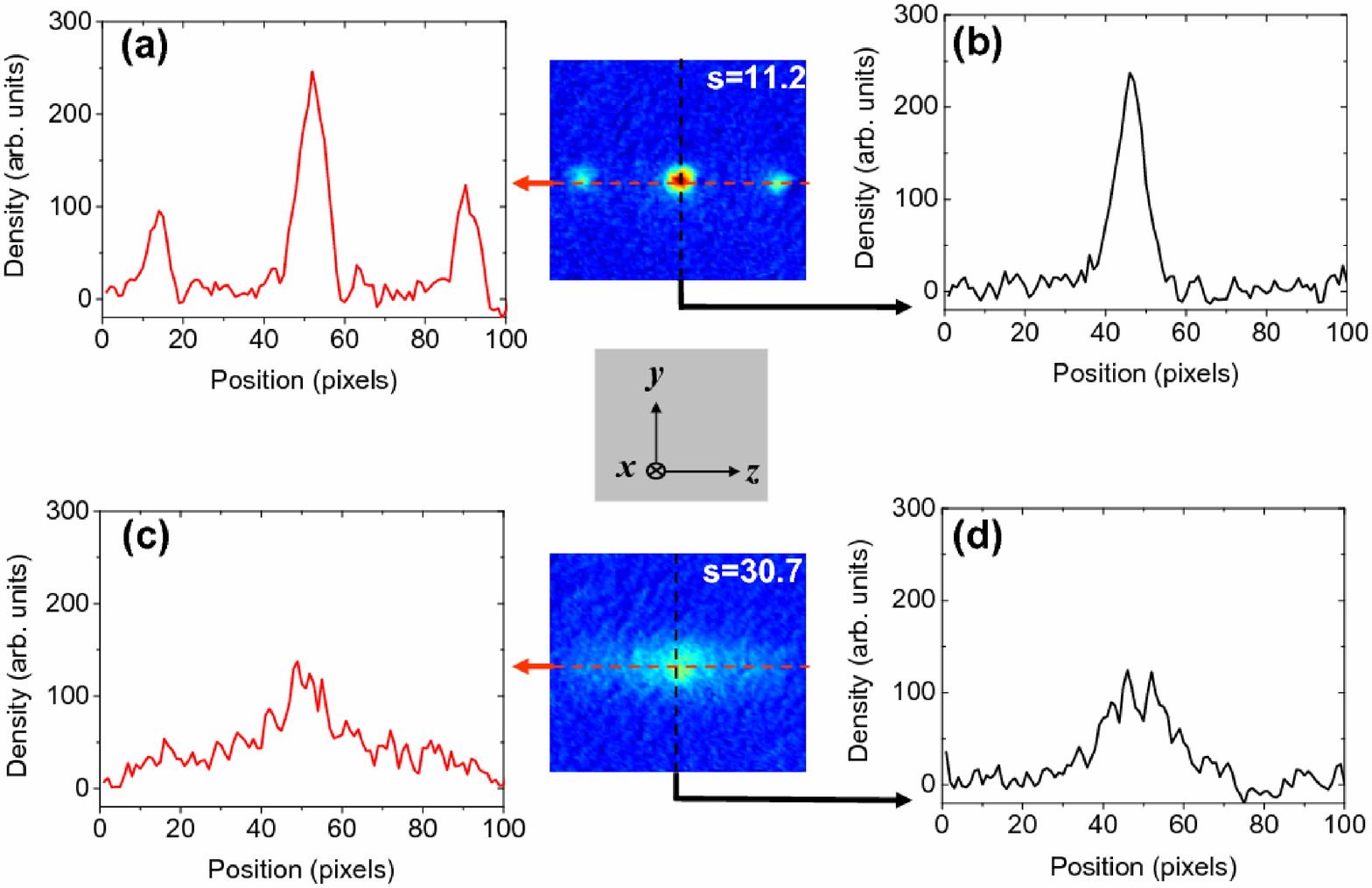}
\caption{Density fluctuations of the released atomic clouds. (\textbf{a})-(%
\textbf{d}) are one-dimensional cuts through the corresponding images. For $%
s=30.7$, the density distribution is highly fluctuated along both the
horizontal ($z$) and vertical ($y$) directions.}
\end{figure}

%\textit{Because there are beyond one hundred lattice sites occupied by the
%cold atomic gas, we cannot directly get the density distribution of
%vortices, even measuring the density distribution along }$z$\textit{\
%direction \cite{BKTexp}.}
To qualitatively analyze the density fluctuations, we consider the following
density probability distribution%
\begin{equation}
P\left( n\right) =\frac{S\left( n-\delta n/2\right) -S\left( n+\delta
n/2\right) }{\delta n\cdot S_{\mathrm{total}}}\,,  \label{defin}
\end{equation}%
where $\delta n$ is the width of a density interval, $S_{\mathrm{total}}$ is
the total area of the region occupied by the atomic gas, while $S\left(
n\right) $ is the area of the region where the density larger than $n$. The
averaged density is $n_{s}=1/S_{\mathrm{total}}$. For convenience, we define
a dimensionless density as $n_{0}=n/n_{s}$. The dimensionless density
probability distribution is then $P_{0}\left( n_{0}\right) =n_{s}P\left(
n\right) $.

The right column of Fig. 2 displays $P_{0}\left( n_{0}\right) $ calculated
from the corresponding density distribution. In each image, the pixel region
chosen for the calculation of $P_{0}\left( n_{0}\right) $ is that occupied
by the cold atoms, as enclosed by the red dashed ellipse. In all our
calculations of $P_{0}\left( n_{0}\right) $, $\delta n_{0}$ is just the
horizontal spacing of discrete data points. Due to the optical noise, $%
P_{0}\left( n_{0}\right) $ can be nonzero even for negative $n_{0}$. This
can be understood from the atomic density formula based on the absorption
imaging signal. Assuming $n_{0b}$ is the minimum negative density for
nonzero $P_{0}\left( n_{0}\right) $, optical noise concentrates in the
region of $n_{0b}<n_{0}<\left\vert n_{0b}\right\vert $ in each $P_{0}\left(
n_{0}\right) $ plot. Thus, $n_{0}>\left\vert n_{0b}\right\vert $ gives the
effective region where the optical noise is negligible and $P_{0}\left(
n_{0}\right) $ reflects the true density probability distribution.

The blue solid lines in Fig. 2(b1)-(b6) represent the following exponential
density probability distribution%
\begin{equation}
P_{0}^{e}=e^{-\left\vert n_{0}\right\vert }.  \label{exp}
\end{equation}%
It is obvious that, with increased $s$ (and hence $U/J$), $P_{0}\left(
n_{0}\right) $ has a tendency to $P_{0}^{e}$. In Fig. 2(b6), we see that $%
P_{0}\left( n_{0}\right) $ agrees well with $P_{0}^{e}$ in the whole
effective region. Despite of the extremely complex many-body state in the QC
region above the BKT transition, our results show a simple universal
behaviour.

We have numerically simulated our experiments to explain the exponential
density probability distribution. When a released gas has experienced a free
expansion over a time of $t$, the 2D density distribution is given by

\begin{widetext}
\begin{equation}
n_{2D}\left( y,z,t\right) =\int dx\left\vert
\sum_{k}\sqrt{N_{k}}e^{i\left( \phi _{k\perp }\left( x,y,t\right)
+\phi _{k}^{s}\left( z,t\right) \right) }\varphi _{k\perp }\left(
x,y,t\right) \varphi _{kz}\left( z,t\right) \right\vert
^{2}+n_{2D}^{opt}\left( x,y\right) .  \label{densitysimu}
\end{equation}
\end{widetext}

Here $\varphi _{k\perp }\left( x,y,t\right) \varphi _{kz}\left( z,t\right) $
is the wave packet of the expanded subcondensate initially in the $k$th
lattice site, with $|\varphi _{k\perp }\left( x,y,t\right) |^{2}$ being the
Thomas-Fermi density distribution in transverse ($x$-$y$) directions, and $%
|\varphi _{kz}\left( z,t\right) |^{2}$ the Gaussian density distribution
along $z$ direction. Two classes of phase fluctuations enter the expression.
$\phi _{k\perp }\left( x,y,t\right) $ represents the spatial phase
fluctuations in transverse directions, while $\phi _{k}^{s}\left( z,t\right)
$ comprises the relative phase fluctuations between different
subcondensates. $N_{k}$ denotes the atomic number of the $k$th
subcondensate. The term $n_{2D}^{opt}$ is added to account for the optical
noise. In numerical calculations, it was simulated with the realistic
optical noise distribution in the region having no atoms.

\begin{figure}[h]
\includegraphics[width=0.8\linewidth,angle=0]{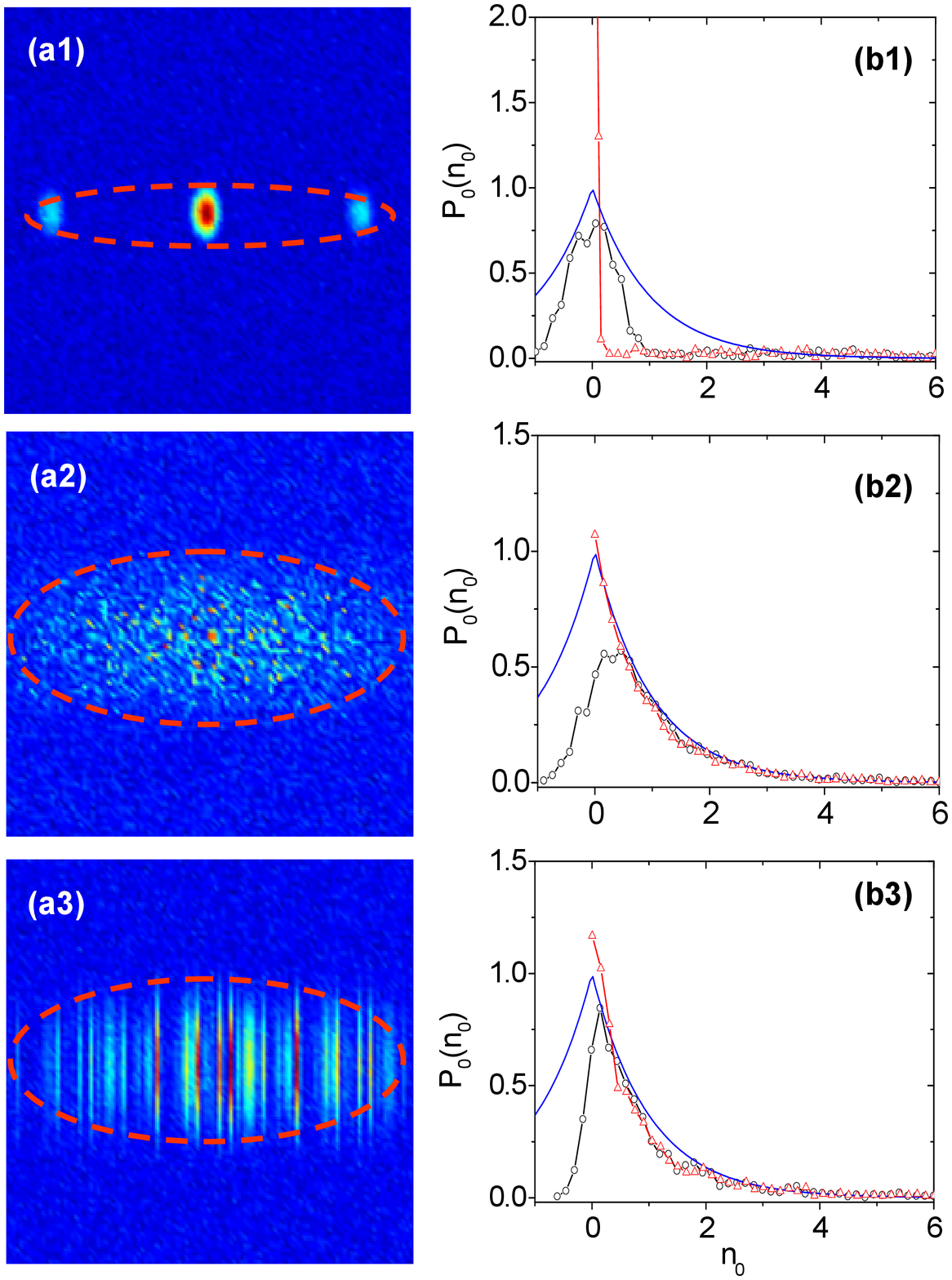}
\caption{Theoretical simulation. Left column: simulated density
distributions based on Eq. \eqref{densitysimu}. (\textbf{a1}), (\textbf{a2})
and (\textbf{a3}) are three typical cases in which an atomic gas initially
stays in SF, Type-I QC and Type-II QC regions, respectively. In (\textbf{a1}%
), $s=11.2$, $\protect\delta \protect\phi _{k\perp }=0.2\protect\pi $. In (%
\textbf{a2}), $s=30.7$, both $\protect\phi _{k}^{s}$ and $\protect\phi %
_{k\perp }$ are completely random. In (\textbf{a3}), $s=30.7$, $\protect\phi %
_{k\perp }=0$, but $\protect\phi _{k}^{s}$ is completely random. Each of the
color map has a size of $0.9\times 0.9$ \textrm{mm}$^{2}$. Right column:
Black lines are the density probability distribution calculated from the
corresponding density distributions in the left column. Red lines are
obtained under the same conditions except that the optical noise is not
included. The exponential density probability distribution (blue lines) is
also plotted for comparison. The data in (\textbf{b2}) clearly shows an
exponential density probability distribution in the QC region above the BKT
transition.}
\end{figure}

Figs. 4(a1) and (b1) give the simulated density distribution and density
probability distribution for an atomic gas initially in the combined trap
with $s=11.2$. Spatial phase fluctuations ($\phi _{k\perp }$) have been
assumed to be zero. Figs. 4(a2) and (b2) display similar calculations for $%
s=30.7$, but with completely random phases $\phi _{k\perp }$ and $\phi
_{k}^{s}$. The theoretical results in Figs. 4(b2) match the exponential
density probability distribution, in agreement with our experimental
results. The coexistence of two random phases of $\phi _{k\perp }$ and $\phi
_{k}^{s}$ makes the system similar to the situation of the interference of
randomly scattered waves \cite{Sheng}, where exponential distribution is
also found. More general analytical derivations for Eq. (\ref{densitysimu})
and the exponential density probability distribution are given in the
Appendix. Figs. 4(a3) and (b3) give the simulation of $s=30.7$ for a cold
atomic gas in the QC region below the BKT transition. Accordingly, $\phi
_{k\perp }$ is assumed to be zero, while there exists a completely random
relative phase in $\phi _{k}^{s}$. No exponential density probability
distribution is found in this simulation. In addition, we do not notice
regular interference fringes either. Note that in Ref. \cite{Had}, the QC
region far below the BKT transition was studied experimentally. However,
because there are only about $30$ subcondensates, high-contrast interference
fringes were still observed. Since our calculation is based on the
experimental parameters in this work, the total number of subcondensates is
much higher (up to $250$).

To further explain the experimental phenomena, we stress that there are two
completely different situations in determining the BKT transition
temperature for cold atoms in 1D optical lattices.

\textbf{(1) BKT transition in QC and MI regions:} In these two regions, the
quasi-2D Bose gases in different lattice sites can be regarded as
independent subcondensates as far as the BKT transition is concerned. By
minimizing the free energy $F=E-TS$ ($E$ and $S$ are the energy and entropy
of the Bose gas in a lattice site when vortices are considered), one can get
the BKT transition temperature $T_{BKT}=T_{2D}/4$ \cite{Pethick}. In the MI
region above the BKT transition, Eq. (\ref{densitysimu}) can be used to
simulate the density distribution by including a completely random relative
phase in $\phi _{k}^{s}$ \cite{Had}, if the averaged particle number per
lattice site is much larger than $1$. Thus, in the MI region above the BKT
transition, one still expects the exponential density probability
distribution for a large number of more than one hundred independent
subcondensates.

\textbf{(2) BKT transition in SF region:} In the SF region, although the
Bose gases in the lattice sites become quasi-2D, they are highly correlated.
For a series of $M$ highly correlated quasi-2D Bose gases, the free energy
becomes $F=ME-TS$. If there are vortices in the quasi-2D subcondensates,
vortices in different lattice sites are highly correlated in spatial
locations. Thus, $S$ can be approximated as the entropy of a single
subcondensate. Then the BKT transition temperature becomes $%
T_{BKT}^{M}=MT_{BKT}$. Since $M>100$, we see that $T_{BKT}^{M}>>T_{2D}$.
This means that the temperatures of quasi-2D condensates are always much
lower than the BKT transition temperature. For the situation of $s=11.2$ in
our experiment, based on the estimation of the particle number fluctuations $%
\delta N_{k}$ \cite{Jav} and the relation of $\delta \phi _{k}^{s}\sim 2\pi
/\delta N_{k}$ \cite{Orzel}, we deduced a phase fluctuation of $\delta \phi
_{k}^{s}\approx 0.2\pi <<2\pi $, which shows strong correlation between
different lattice sites. As shown in Fig. 3(b), there are no noticeable
density fluctuations along $y$ direction, which confirms that the initially
confined gas is well below the BKT transition even though $T$ is already
beyond $T_{BKT}$.

We should also emphasize that, in the QC region above the BKT transition,
the system cannot be regarded as a classical thermal cloud although the
temperature is higher than the critical temperature $T_{c}$. A classical
thermal cloud features density fluctuations as $\delta ^{2}n\approx n$. In
contrast, the exponential density probability distribution corresponds to
much larger density fluctuations of $\delta ^{2}n\approx n^{2}$ \cite%
{XiongWu}. This exponential density probability distribution physically
originates from the superposition of quantum sates with a completely random
phase, while the universal behaviour lies in that the distribution is always
exponential for completely random superposition of quantum state \cite%
{XiongWu}. The present work supports the long-standing belief that the
quantum states in the QC region are strongly correlated, and there should be
a simple universal behaviour, once appropriate physical quantity is found
\cite{Sachdev,Cole,Gegen}.

\begin{acknowledgments}
H. W. thank the cooperation with Biao Wu about the dynamical
universal behaviour of quantum chaos, which stimulates the present
work. This work was supported by National Key Basic Research and
Development Program of China under Grant No. 2006CB921406 and NSFC
under Grant No. 10634060.
\end{acknowledgments}

\begin{center}
\textbf{Appendix}
\end{center}

\noindent \textbf{Bose-Hubbard model.} The combined potential of a
one-dimensional (1D) optical lattice and a harmonic trap is given by
\begin{subequations}
\begin{equation}
V=\frac{1}{2}m\omega _{\perp }^{2}\left( x^{2}+y^{2}\right) +\frac{1}{2}%
m\omega _{z}^{2}z^{2}+sE_{R}\sin ^{2}\left( \frac{2\pi z}{\lambda }\right) .
\label{potential}
\end{equation}%
Here $E_{R}$ is the recoil energy of an atom absorbing one lattice photon
with wavelength $\lambda $. For an atom in a lattice site, it experiences an
effective harmonic potential along $z$ direction with angular frequency $%
\widetilde{\omega }_{z}\simeq 2\sqrt{s}E_{R}/\hbar $.

Bose-condensed gases in 1D, 2D and 3D optical lattices are widely studied by
the following Bose-Hubbard model \cite{FisherS}
\begin{equation}
\widehat{H}=-\sum\limits_{\left\langle ij\right\rangle }J\widehat{a}%
_{i}^{\dag }\widehat{a}_{j}+\frac{U}{2}\sum\limits_{i}\widehat{n}_{i}\left(
\widehat{n}_{i}-1\right) +\sum\limits_{i}\left( \varepsilon _{i}-\mu \right)
\widehat{n}_{i}.  \label{BH}
\end{equation}%
Here $\widehat{n}_{i}=\widehat{a}_{i}^{\dag }\widehat{a}_{i}$ is the number
operator at the $i$th site, with $\widehat{a}_{i}^{\dag }$ ($\widehat{a}%
_{i}) $ being the boson creation (annihilation) operator. $U$ and $J$
represent the strengths of the on-site repulsion and of the nearest neighbor
hopping, respectively. $\varepsilon _{i}$ describes an energy offset due to
the harmonic trap, and $\mu $ is the chemical potential. The phase diagram
given by Fig. 1 in the text is calculated with the above combined potential
and Bose-Hubbard model.

\noindent \textbf{Universal exponential density probability distribution.}
The many-body quantum state in the QC region can be written as in a general
way

\begin{equation}
\left\vert \Psi _{qcr}\right\rangle =\sum_{\left\{ \Sigma N_{k}=N\right\}
}C\left( N_{1},\cdot \cdot \cdot ,N_{2k_{M}+1}\right) \prod_{k}\frac{1}{%
\sqrt{N_{k}!}}\left( \widehat{a}_{k}^{\dag }\right) ^{N_{k}}\left\vert
0\right\rangle .
\end{equation}%
Here $N_{k}$ denotes the atomic number in the $k$th lattice site. The total
atomic number is then $N=\Sigma _{k=-k_{M}}^{k=k_{M}}N_{k}$, with $2k_{M}+1$
being the total number of occupied lattice sites. For strongly correlated
quantum gases in the QC region, the function $C\left( N_{1},\cdot \cdot
\cdot ,N_{2k_{M}+1}\right) $ is extremely complex, and no analytical
expression exists. After switching off the combined potential, the density
distribution is then%
\begin{equation}
n\left( x,y,z,t\right) =\left\langle \Psi _{qcr}\left( t\right) \right\vert
\widehat{\Psi }^{\dag }\widehat{\Psi }\left\vert \Psi _{qcr}\left( t\right)
\right\rangle .
\end{equation}

Assuming that the initial wave function in the $k$th lattice site is $%
\varphi _{k}\left( x,y,z,t=0\right) =\varphi _{k\perp }\left( x,y,t=0\right)
\varphi _{kz}\left( z,t=0\right) $, for long-time evolution so that the
expanded subcondensates have sufficient overlapping with each other, we have%
\begin{equation}
n\left( x,y,z,t\right) \approx \left\vert \sum_{k}\sqrt{N_{k}}e^{i\phi
_{k}^{s}\left( z,t\right) }\varphi _{k\perp }\left( x,y,t\right) \right\vert
^{2}\left\vert \varphi _{0z}\left( z,t\right) \right\vert ^{2}.
\label{densitysum}
\end{equation}%
This formula holds in the QC region, because there are non-negligible
particle number fluctuations in each lattice site.\textit{\ }Here $\phi
_{k}^{s}\left( z,t\right) =\phi _{k}^{r}+\phi _{k}^{0}$, with $\phi _{k}^{r}$
being a completely random relative phase in the QC region. $\phi
_{k}^{0}=m\left( z-k\lambda /2\right) ^{2}/2\hbar t$\ can be obtained
directly from the free expansion along $z$\ direction. Without the random
phase $\phi _{k}^{r}$, the phase $\phi _{k}^{0}$ will lead to clear
interference fringes along $z$ direction with a period of $4\pi \hbar
t/m\lambda $ \cite{PedriS}. It is understood that there would be no
interference fringes along $z$ direction if $\phi _{k}^{s}$ is a completely
random relative phase, as shown in Figs. 2(a5)-(a6) in the text.

In the QC region above the BKT transition, there are two sorts of random
phases: (i) the completely random phase $\phi _{k}^{s}\left( z,t\right) $ in
the QC region; (ii) the spatially relevant random phase $\phi _{k\perp
}\left( x,y,t\right) $ above the BKT transition. In this region, the density
distribution can be further written as

\begin{eqnarray}
n\left( x,y,z,t\right) &\approx &\left\vert \sum_{k}\sqrt{N_{k}}e^{i\phi
_{k\perp }\left( x,y,t\right) }e^{i\phi _{k}^{s}\left( z,t\right)
}\right\vert ^{2}  \notag \\
&&\left\vert \varphi _{0\perp }\left( x,y,t\right) \varphi _{0z}\left(
z,t\right) \right\vert ^{2}.
\end{eqnarray}%
Here $\phi _{k\perp }\left( x,y,t\right) $ physically originates from the
spatial phase fluctuations of quasi-2D Bose gas above the BKT transition.

The density distribution recorded by a CCD is a 2D density distribution $%
n_{2D}\left( y,z,t\right) =\int dxn\left( x,y,z,t\right) $. By assuming that
$\left\vert \varphi _{0\perp }\left( x,y,t\right) \varphi _{0z}\left(
z,t\right) \right\vert ^{2}$ is locally constant (i.e. the local density
approximation), after integrating out the $x$ variable, we have%
\begin{eqnarray}
n_{2D}\left( y,z,t\right) &\approx &\left\vert \frac{1}{\sqrt{2k_{M}+1}}%
\sum_{k}\alpha _{k}\left( y,z,t\right) e^{i\beta _{k}\left( y,z,t\right)
}\right\vert ^{2}  \notag \\
&&n_{2D}^{0}\left( y,z,t\right) ,
\end{eqnarray}%
with $n_{2D}^{0}\left( y,z,t\right) =\int dx\left\vert \sqrt{N}\varphi
_{0\perp }\left( x,y,t\right) \varphi _{0z}\left( z,t\right) \right\vert
^{2} $. $\alpha _{k}\left( y,z,t\right) $ and $\beta _{k}\left( y,z,t\right)
$ are statistically independent random real function, because they
originates from two different random phases $\phi _{k}^{s}\left( z,t\right) $
and $\phi _{k\perp }\left( x,y,t\right) $. $\beta _{k}\left( y,z,t\right) $
distributes uniformly between $-\pi $ and $\pi $. The normalization
condition requests $\left\langle \alpha _{k}^{2}\right\rangle =1$.

Based on the central limit theorem \cite{Feller}, we get the following
exponential density probability distribution%
\begin{equation}
P\left( n_{2D}\left( x,y,t\right) \right) =\frac{e^{-n_{2D}\left(
x,y,t\right) /n_{2D}^{0}\left( x,y,t\right) }}{n_{2D}^{0}\left( x,y,t\right)
}.  \label{probabilityx-y}
\end{equation}

For uniform $n_{2D}^{0}$, it is straightforward to get Eq. (2) in the text.
For nonuniform $n_{2D}^{0}$, with the local density approximation, Eq. (2)
in the text is a very good approximation obtained from Eq. (\ref%
{probabilityx-y}).

In Figs. 2(b5)-(b6) in the text, we see that the density probability
distribution agrees well with the exponential density probability
distribution. However, because the system temperature is slightly smaller
than the critical temperature for Fig. 2(b5) ($s=22.3$), we expect Fig.
2(b6) agrees better with the exponential density probability distribution.
Here we give the semilog plot of the density probability distribution for $%
s=22.3$ and $s=30.7$, respectively. We do find that the situation of $s=30.7$
agrees better with the exponential density probability distribution.

\begin{figure}[h]
\centering
\includegraphics[width=0.6\columnwidth,angle=270]{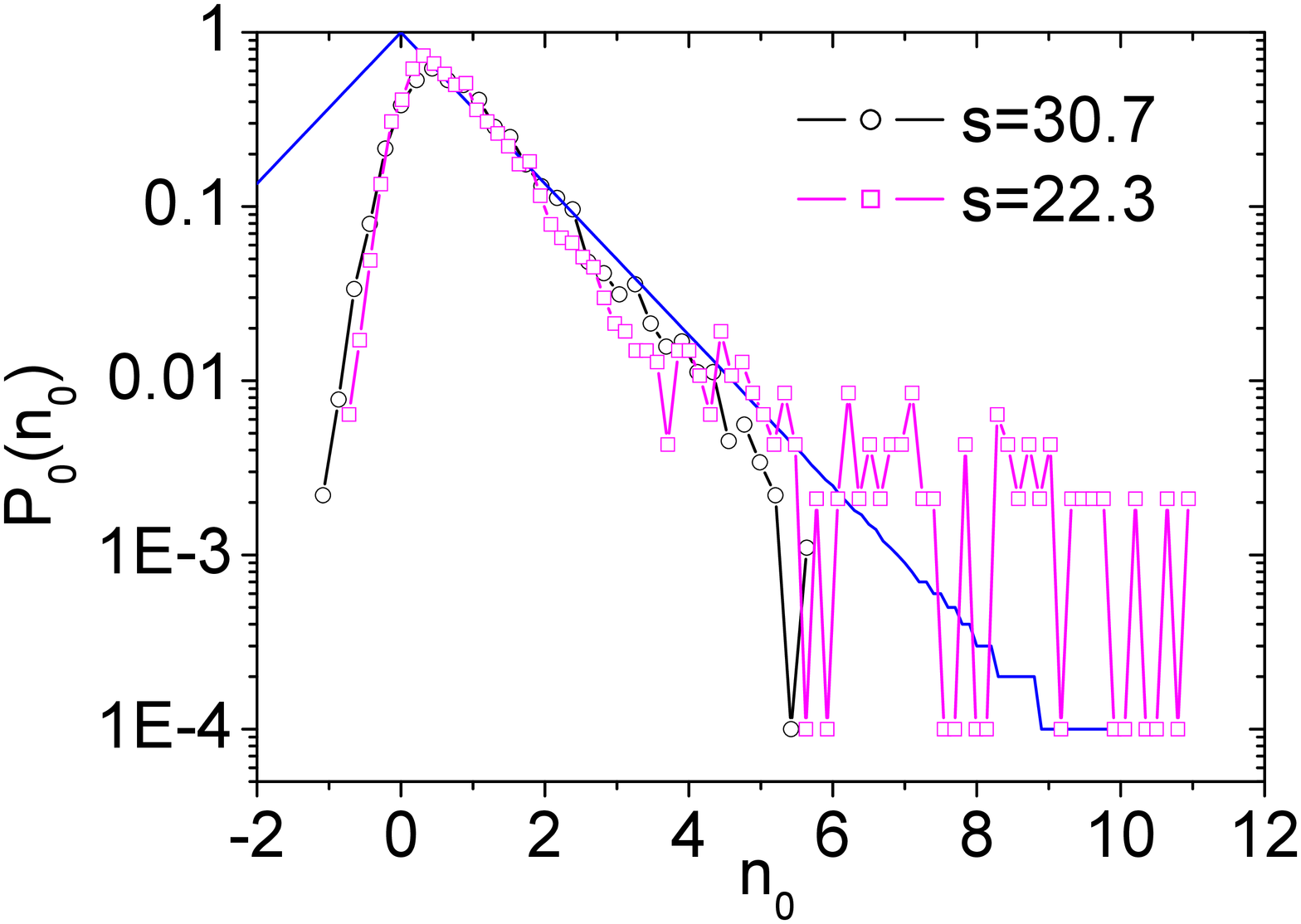}
\caption{Semilog plot of the density probability distribution for $s=22.3$
and $s=30.7$. The blue line gives the exponential density probability
distribution.}
\end{figure}

\noindent \textbf{System temperature.} In our experiment, the total atomic
number can be accurately measured. The atomic numbers of Figs. 2(a4)-(a6) in
the text are $1.10\times 10^{5}$, $1.19\times 10^{5}$ and $1.13\times 10^{5}$%
, respectively. To drive the system into the deeper region of superfluidity,
we have to further lower the temperature of an atomic sample before loading
it to the lattice. This is implemented by kicking more atoms out of the
magnetic trap during the evaporative cooling stage. Specifically, for Figs.
2(a2) and (a3) in the text, the atomic numbers are reduced by about a factor
of two compared with Figs. 2(a4)-(a6) in the text.

However, the measurement of the system temperature is hampered by the lack
of accurate thermometry in optical lattices. Actually, the ratio $T/T_{c}$
is a more important quantity than $T$ in characterizing the transition from
superfluid to quantum critical regions. Therefore, the inaccuracy of system
temperature does not affect the physics of the present work, as long as $%
T/T_{c}$ can be somehow determined with a satisfying accuracy.

As predicted in Ref. \cite{Duan} and demonstrated in a recent experiment
\cite{Phillips}, the appearance of superfluid in an inhomogeneous lattice is
associated with a bimodal structure in the density profile. The condensate
fraction $N_{0}/N$ can be obtained by fits to the bimodal distribution. As
we have known, $N_{0}/N$ shows a temperature dependence having a
characteristic shape as
\begin{equation}
N_{0}/N=1-(T/T_{c})^{\alpha }.  \label{SF-fraction}
\end{equation}%
Apparently, $T/T_{c}$ can be directly determined if the parameter $\alpha $
is known. Blakie \emph{et al.} \cite{Tc-Blakie} had performed numerical
calculations of the critical temperature as well as the condensate fraction,
for a combined trap similar to that in our work. Their results show that $%
\alpha $ is close to $3$ at a moderate lattice depth, but it may drop
slightly with increasing lattice depth. Nevertheless, at a high depth level,
the condensate fraction is usually very small, and, as a result, the value
of $T/T_{c}$ derived from Eq. \eqref{SF-fraction} becomes insensitive to the
potential errors of $\alpha $. Therefore, we just set $\alpha =3$ in the
calculation of $T/T_{c}$, which seems to be a reasonable assumption.

\begin{figure}[h]
\centering
\includegraphics[width=0.6\columnwidth,angle=270]{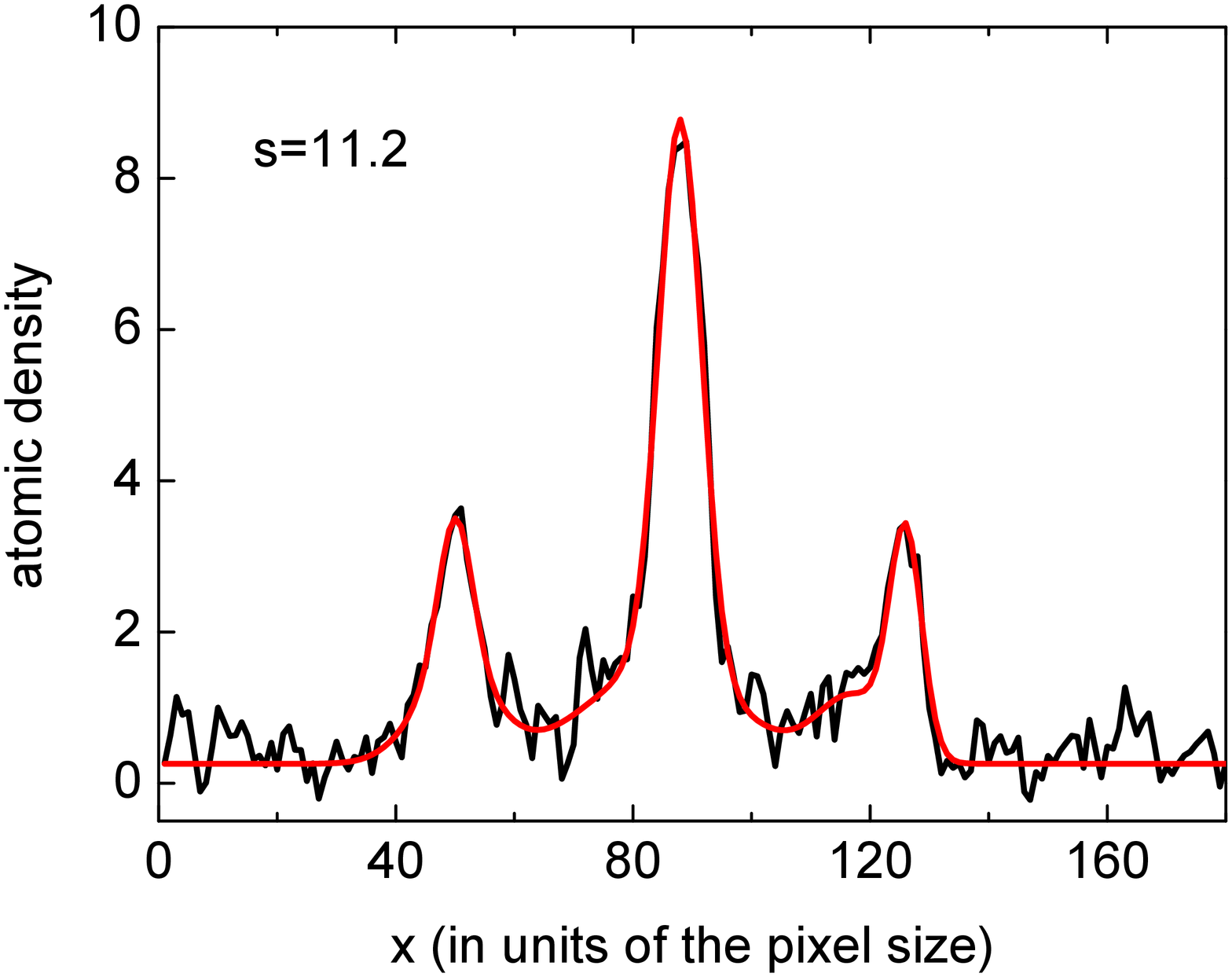}
\caption{A bimodal fit to the interference pattern in Fig. 2(a3) in the
text. The total atom number is measured as $N\simeq 5.9\times 10^{4}$, and
the atomic cloud was released from the combined trap with a lattice strength
of $s=11.2$. Black line, the measured atomic density which is obtained by
integrating over each column of pixels. Red line, a fitting curve including
six Gaussian peaks, where the three narrow peaks at the upper part are
attributed to superfluid. The condensate fraction is $N_{0}/N\approx 0.58$,
corresponding to $T/T_{c}\approx 0.75$.}
\end{figure}

For each absorption image, we first obtain the atomic density along the $z$
dimension by integrating pixels in each column. The narrow peaks riding over
broad ones are regarded as the condensate parts. We then fit both the broad
and the narrow peaks using Gaussian profiles, so as to get the condensate
fraction. Apparently, this method works only when $T<T_{c}$. In Fig. 2(a6)
in the text, the interference peaks vanish completely, indicating that $%
T>T_{c}$. In this case, we are unable to extract the quantity $T/T_{c}$ any
more with this method. The universal exponential density probability
distribution found in this work also implies that the accurate measurement
of system temperature in the QC region above BKT transition is difficult.

\end{subequations}

\end{document}